\documentclass[12pt]{article}
\usepackage[applemac]{inputenc}
\usepackage[margin=2.5cm]{geometry}
\usepackage{amssymb,amsmath}
\usepackage{graphicx,color}
\usepackage[rm,small]{caption}
\usepackage{cite,url}
\usepackage[T1]{fontenc}
\usepackage{lmodern}
\usepackage{mathrsfs}
\usepackage{hyperref}

\newcommand{\kB}{\ensuremath{k_{\mathrm{B}}}} 
\newcommand{\ave}[1]{\ensuremath{\left \langle {#1} \right \rangle}} 
\newcommand{\eps}{\ensuremath{\varepsilon}} 

\newcommand{\om}{\ensuremath{\omega}}

\newcommand{\C}{\ensuremath{\mathcal{C}}}

\newcommand{  \p   }   { \ensuremath{      \big{(}          }               }
\newcommand{  \q   }   {  \ensuremath{      \big{)}        }                }
\newcommand{\g}[1]{``{#1}''}

\title{On Classical Gases.}
\author{%
Jacques Arnaud\\
\normalsize{\itshape Mas Liron, F30440 Saint Martial, France}\\
Laurent Chusseau\\
\normalsize{\itshape IES, UMR n°5214 au CNRS, Université Montpellier II, F34095 Montpellier, France}\\
Fabrice Philippe\\
\normalsize{\itshape LIRMM, UMR n°5506 au CNRS, 161 rue Ada, F34392 Montpellier, France}%
}
\date{\today}

\begin{document}

\maketitle

\begin{abstract}

The ideal gas laws are derived from the democritian concept of corpuscles moving in vacuum plus a principle of simplicity, namely that these laws are independent of the laws of motion aside from the law of energy conservation. A single corpuscle in contact with a heat bath and submitted to a $z$ and $t$-invariant force $-w$ is considered, in which case corpuscle distinguishability is irrelevant. The non-relativistic approximation is made only in examples. Some of the end results are known but the method appears to be novel. The mathematics being elementary the present paper should facilitate the understanding of the ideal-gas law and more generally of classical thermodynamics. It supplements importantly a previously published paper: The stability of ideal gases is proven from the expressions obtained for the force exerted by the corpuscle on the two end pistons of a cylinder, and the internal energy. We evaluate the entropy increase that occurs when the wall separating two cylinders is removed and show that the entropy remains the same when the separation is restored. The entropy increment may be defined at the ratio of heat entering into the system and temperature when the number of corpuscles (0 or 1) is fixed. In general the entropy is defined as the average value of $\ln(p)$ where $p$ denotes the probability of a given state. Generalization to $z$-dependent weights, or equivalently to arbitrary static potentials, is made.

\end{abstract}


\section{Introduction}\label{introduction}

This paper gives an alternative derivation of the classical barometric and ideal gas laws. In the title the word ``classical'' means: ``non-quantum'' ($\hbar\to 0$). Our results coincide with those obtained from the Bohr-Sommerfeld (BS) quasi-classical approximation of Quantum Mechanics and the Boltzmann factor but the method is more straightforward. Initially we obtained the average force $\ave {F}$ exerted by a corpuscle on a piston from the BS theory. The corpuscle action (area in the position-momentum $z$-$p$ phase-space, where $z$ denotes the corpuscle altitude and the momentum $p=-w\,t$, where $-w$ is the force applied to the corpuscle, e.g., its weight, and $t$ is time) is discrete, and evenly spaced in units of the Planck constant $2\pi\hbar$. Because the ``bouncing ball'' presently considered is not an harmonic oscillator the discrete corpuscle energies are not evenly spaced. This is why, when going to the continuous limit converting the sum into an integral we must introduce a distribution $\omega(z_m)$ in the form given later. After going through these semi-classical considerations we discovered that it was sufficient to postulate the simplicity principle according to which the average force must not depend on the equations of motion. This concept, unlike the quantum theory, could have been understood at the time of the ancient Greece. Going the opposite way, one may say that the simplicity principle just stated suggests the semi-classical quantum theory. The entropy should be defined in general as the average value of $\ln(p)$ where $p$ denotes the probability of a state, which accounts for a possible uncertainty concerning the presence of a corpuscle.

By \emph{ideal gas}, we mean a collection of non-interacting corpuscles. We may therefore restrict ourselves to a single corpuscle so that considerations of corpuscles distinguishability are irrelevant. We suppose that we know with certainty whether a corpuscle is present or not, except in Section \ref{separation}. Photons whose number may vary do not fulfill our definition of an ideal gas even though they are non-interacting; thus they are not treated. 

Only motion along the vertical $z$ axis is considered, but generalization of the barometric and ideal-gas law to three dimensions is straightforward. As said earlier, the corpuscle is submitted to an external force $-w$, perhaps of electrical origin, $w$ being called the corpuscle \g{weight}, or equivalently to a potential $\Phi(z)=w\,z$. Because the presence of a corpuscle affects negligibly the potential the latter is an \emph{external} potential. We call \emph{perfect gas} ( ``gaz parfait'' in french, see\cite{Conche:2002}. We quote: ``Le désordre initial de Démocrite peut faire songer au chaos moléculaire de Boltzmann qui, dans sa théorie des gaz parfaits, admettait comme une simple évidence que dans la situation de départ positions et vitesses des molécules sont réparties au hasard, avec l’hypothèse générale d’indépendance de ces paramètres deux à deux, indépendance que l’on trouve aussi chez Démocrite'') an ideal gas with no external force acting on the corpuscle except at the boundaries and with the non-relativistic approximation being made. In that case the constant-volume heat capacity $C$ is independent of volume and temperature. These definitions essentially agree with the ones given in\cite{idealgas}. We quote: ``a distinction is made between an ideal gas, where  the heat capacities could vary with temperature, and a perfect gas for which this is not the case''. In our one-dimensional model neglecting corpuscle rotations and vibrations the heat capacity $C$ (derivative of the gas internal energy $U$ with respect to temperature $\theta$ at fixed volume) is equal to 1/2. According to our definition an ideal gas is not necessarily a \g{Joule's gas}: the heat capacity $C$ may depend on temperature $\theta$ and volume $h$. The free expansion of an ideal gas may therefore entail temperature changes while this is not the case for a perfect gas. There is full agreement between our general results and the first and second laws of thermodynamics as they are spelled out in textbooks. There is also full agreement between our results and the usual perfect-gas laws that can be found in the Bernoulli work and elementary textbooks in the limit considered. We give here a generalized and simplified treatment that accounts for the effect of a constant force acting on the corpuscle and (among others) special-relativity effects. Some results in those cases are known, see Landsberg \cite{landsberg} and Louis-Martinez \cite{Louis-Martinez:2011}. 

The present paper is a generalization of our previous papers\cite{Arnaud:2013,arnaud2} where our motivation is explained in more detail than in the present paper. Presently, we calculate the average forces exerted on both ends of a vertical cylinder (they are different when the corpuscle has weight), discuss the gas stability, the entropy increments that occur when a separating wall is removed and restored, and consider weights that may vary with altitude, as is the case for example on earth for cylinder heights that are not negligible compared to the earth radius. 

We first consider the round-trip time $\tau(z_m)$ needed for a corpuscle thrown upward with energy is $E=w\,z_m$ to reach an altitude $z_m$ above the ground level $z=0$ and come back to the ground level. If the corpuscle bounces elastically on the ground, $\tau(z_m)$ represents the oscillation period. We consider only round-trip times, that is time delays measured at some altitude, so that no problem of clock synchronisation arises. The time during which the corpuscle is located above some altitude $z\le z_m$ during a period is: $\tau(\zeta)\equiv\tau(z_m-z)$, since, under our assumption of a constant weight the $\tau$-function does not depend on the initial altitude or initial time. If the weight is of gravitational origin with acceleration $g$, it is known from the Galileo experiment that the function $\tau(\zeta)$ does not depend on the corpuscle mass $m=w/g$ but this is not so in general.

We consider only thermal-equilibrium situations: If we wait a sufficiently long period of time an isolated system ceases to evolve. We take it as an empirical result that, leaving aside general-relativity effects\footnote{According to \emph{general relativity}, thermal energy has weight. But this (so-called Tolman) effect that entails a temperature variation at equilibrium: $\theta(z)=\theta(0)/(1+2z\,g/c^2)$, where $g$ denotes the gravity acceleration and $c$ the speed of light, is entirely negligible; see for example equation (1) of\cite{2013arXiv1302.0724H}. This paper gives the following interpretation of equilibrium: we quote ``The temperature $\theta$ is essentially equal to $\hbar$ divided by the time required by the system to move from one state to the next''. This interpretation leads to the condition of temperature uniformity for weak gravity since in that case time intervals do not depend significantly on altitude.}, two bodies left in contact for a sufficient period of time with energy being allowed to flow from one to the other, reach an equilibrium state corresponding to equal temperatures as one can judge by our senses. If energy may flow spontaneously (without work expenditure) from one body to the other, the converse is not possible: the process is non-reversible (zeroth law of thermodynamics). 

The purpose of the present paper is thus to show that the thermodynamics of ideal gases and particularly the barometric and ideal-gas laws may be obtained on the sole basis of the Democritus model according to which nature consists of corpuscles moving in a vacuum, plus a principle of simplicity: namely that these fundamental laws are \emph{independent} of the law of corpuscle motion: non-relativistic, special relativistic, or otherwise. They apply for example to deep-sea or capillary wave packets (see Appendix \ref{hamilton}). To wit, writing the corpuscle Hamiltonian as $\mathcal{H}(p)+w\,z$, the barometric and ideal-gas laws \emph{do not depend} on the $\mathcal{H}(p)$ function.

In our discussion the temperature $\theta\equiv \kB T$, where $\kB$ denotes the Boltzmann constant and $T$ the temperature in kelvin, enters solely for dimensional reasons. We later show that, remarkably, our expressions of the gas internal energy and force (or pressure) derive from the partial derivatives of the Helmholtz potential (or free energy) $A(\theta,h)$. Some of the subsequent calculations are conventional. The heat delivered by the gas is: $\delta Q=\theta \, dS$, an expression for the entropy $S(\theta,h)$ being given. This result enables us to prove that the formally-introduced temperature $\theta$ \emph{is} a thermodynamic temperature. Indeed, we recover for ideal gases the general Carnot result asserting that the efficiency of reversible thermal engines is: $1-\theta_c/\theta_h$, where $\theta_c$ denotes the cold bath temperature and $\theta_h$ the hot bath temperature. Since Kelvin time this expression \emph{defines} the thermodynamic temperature to within an arbitrary proportionality factor, which is fixed by specifying that $\theta\approx370\times10^{-23}$ joules at the water triple point, that is: $\theta=273.16\, \kB$ if we take $\kB=1.38066... 10^{-23}$ joules as an energy unit. An alternative convention will be suggested.

As said earlier we consider a single corpuscle. Because of the slight thermal motion of the container wall there is an exchange of energy between the corpuscle and the heat bath so that the corpuscle energy slowly varies in the course of time. What we are looking for are averages over arbitrary long time intervals. We recall \cite{callen} that the fundamental Helmholtz relation: $A=A(\theta,h)$ and the fundamental entropy relation: $S=S(U,h)$ have the same mathematical and physical contents, being related by Legendre transforms. We verify that our expression for $A$ leads to stable equilibria. The usefulness of the entropy concept is of course that it may not decrease when constraints inside a thermally isolated object are removed or added (second law of thermodynamics). We will verify that this is so from our model formulas. Finally, we shall consider the case where the weight instead of being a constant varies by steps along the vertical axis, that is, in the limit, the case of arbitrary potentials $\Phi(z)$.

The reader may feel that our statement that an invariance principle \emph{implies} the barometric and ideal-gas laws without anything else is surprising. On the other hand a quick reading of standard books on Thermodynamics may lead other readers to believe that this is a well-known fact. For example, Callen \cite {callen} states correctly that: \g{The essence of the ideal-gas law is that molecules of the gas do not interact. This simple fact \emph{implies} that $\textsf{P}\,\textsf{V}\propto N\,T$}. However, in order to reach this conclusion, that author needs postulate the Newtonian law of motion, quantum theory, and the Boltzmann factor. The claim that corpuscle independence entails the barometric and ideal-gas laws without (almost) anything else therefore does not appear to have been justified before.

\begin{figure}
\centering
\includegraphics[width=0.5\columnwidth]{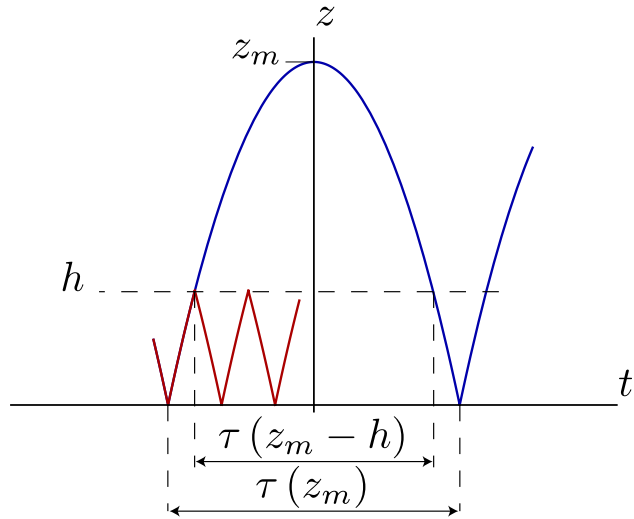}
\caption{Space-time ($z,t$) trajectory for a corpuscle of weight $w$ bouncing off the ground ($z$=0). The maximum altitude reached by the corpuscle is $z_m=E/w$, where $E$ denotes the energy. The motion is periodic with period $\tau(z_m)$, where $\tau(\zeta)$ denotes the corpuscle round-trip time at a distance $\zeta$ from the top of the trajectory. When the maximum altitude is restricted to $h$ by a piston (dashed horizontal line) the motion remains periodic with a period evidently equal to: $\tau(z_m)-\tau(z_m-h)$. Note that this expression holds even if the motion is not symmetric in time.}
\label{figure}
\end{figure}

\section{The barometric law}\label{barometric}

We consider a single corpuscle moving only along the vertical $z$ coordinate and submitted to a constant force $-w$. If the corpuscle energy is $E$, the maximum altitude reached is $z_m=E/w$. The round-trip time $\tau(z_m)$ is the motion period, and the time per period during which the corpuscle is above $z$ is $\tau(z_m-z)$. If follows that the fraction of time during which the corpuscle is above the $z$-level is $\tau(z_m-z)/\tau(z_m)$. To obtain an expression independent of the $\tau(.)$-function, one must introduce an energy distribution $\om(E)$, and define the average time as:
\begin{align}\label{baro}
\ave{above~z}=\frac{\int_{z}^\infty dz_m\, \om(z_m)\tau(z_m-z)/\tau(z_m)}{\int_0^{\infty}\, dz_m\,  \om(z_m)}.
\end{align}
The desired result is obtained for the distribution: $\om(E)=\exp(-w z_m/\theta)\tau(z_m) $:
\begin{align}\label{bar}
\ave{above~z}=\frac{\int_{z}^\infty dz_m\, \exp(-w z_m/\theta)\tau(z_m-z)}{\int_0^{\infty}\, dz_m\,  \exp(-w z_m/\theta)\tau(z_m)}=\exp(-w z/\theta).
\end{align}
In the above integrals going from $h$ to $\infty$ we have replaced $\exp(-w z_m/\theta)$ by $\exp(-w z/\theta)\\
\exp(-w (z_m-z)/\theta)$ and introduced the variable $\zeta\equiv z_m-z$, so that all the integrals go from zero to infinity and cancel out. Note that even though integral signs have been introduced no integration has been performed. We simply consider an integral as a sum of terms and employ the rule of addition associativity. We have employed also the fact that $f(x)=\exp(x)$ is the only function such that $f(a+b)=f(a)f(b)$. Here $\theta$ is an energy introduced on the sole basis of dimension, later on proven to be a thermodynamic temperature. Thus a general form of the barometric law has been obtained without invoking the Boltzmann distribution. For a three-dimensional configuration corresponding to real atmospheres the same result holds: The $\tau(.)$-function is modified, but the barometric law does not depend on that function and therefore the result in \eqref{baro} is unaffected. The more complicated case where $w$ depends on $z$ is treated at the end of the paper in Section \ref{barome}. The result amounts to replacing $w\,z$ in the above expression by the potential $\Phi(z)$. This known result is obtained here most simply. The average forces exerted by the corpuscle on the cylinder lower and upper pistons are evaluated using a similar method in the following section \ref{force}. 

\section{Average force exerted by a corpuscle on pistons}\label{force}

We consider a unit-area cylinder with vertical $z$-axis at some temperature. The bottom of the cylinder is located at altitude $h_o$ and a tight piston at altitude $h_1=h_o+h$ is free to move in the vertical direction. The cylinder contains a single corpuscle submitted to a force $-w$ constant in space and time. For convenience we call $w$ the \g{weight} even though the force may not be of gravitational origin. The corpuscle motion may be relativistic or not. We evaluate the average force $\ave{F_o}$ exerted by the corpuscle on the cylinder bottom and the average force $\ave{F}$ exerted by the corpuscle on the piston. The following relation necessarily holds: $\ave{F_o}+\ave{F}+w=0$. When $w=0$ the forces exerted on both ends are opposite: $\ave{F_o}+\ave{F}=0$. 

For the sake of comparison with textbook formulas note that in our one-dimensional model the average force $\ave{F}$ corresponds to the pressure $\textsf{P}$, the height $h$ corresponds to the volume $\textsf{V}$, and $N=1$. Then, in the non-relativistic approximation, the perfect-gas formula for a mole ($N_A$ corpuscles): $\textsf{P}\,\textsf{V}=R\,T$ where $R=\kB \,N_A$, reads: $\ave{F}\,h=\theta$ where $\theta=\kB T$. Our result provides the ideal-gas law in a generalized form taking into account the weight $w$. In that case, the force at the bottom of the cylinder exceeds in absolute value the force on the piston. We do not discuss the forces that the corpuscle would exert on the cylinder walls in a three-dimensional geometry.

\paragraph{Perfect gases:}

Since $w=0$ we have: $-F_o=F$. Let us recall the textbook result, and explain why it is not entirely satisfactory. The mechanical force exerted by the corpuscle on the piston is equal to the change $2p$ of corpuscle momentum times the number of collisions per unit time $F=2p/\tau$, where the motion period: $\tau=2h/v$ if the corpuscle speed is $v$. Thus $F\,h=v\,p$, where  $v\,p$ is some function of the energy $E$: $v\,p=f(E)$. We obtain the average force for a normalized energy distribution $\om(E)$ as: $\ave{F}h=\ave{vp}\equiv\int_0^\infty dE\,\om(E)\,f(E)$. To obtain a definite solution one usually makes the non-relativistic approximation $v=p/m$ and thus $v\,p=p^2/m=2E$, where $m$ is the corpuscle mass, using the expression of the kinetic energy to which $E$ reduces in the present case. We obtain the textbook result: $\ave{F}h=2\ave{E}\equiv 2U$. The quantity on the right-hand side is identified with the temperature $\theta\equiv \kB T$ and we finally obtain: $\ave{F}h=\theta$. This is the perfect gas law (often referred to as the \g{ideal gas law}) for a single corpuscle. However, the definition of $\theta$ given above is to a large degree arbitrary. For example, consider the ultra-relativistic case. Then $v=c$ where $c$ denotes the speed of light and $v\,p=E$. This forces us to redefine the temperature $\theta$ as being equal to $U$ instead of $2U$. In other words, the conventional formulation does not enable us to obtain the perfect gas law in a general form. Furthermore, the average value $U$ of $E$ depends on the distribution $\om(E)$, which is presently unknown. 

Alternatively one could call \emph{temperature} the force that must be applied on the piston of a cylinder containing a gas such as helium to maintain $h$ at some fixed value. Then the empirical Boyle-Mariotte is established, and the Gay-Lussac law that asserts that $\ave{F}h$ is proportional to temperature follows from the definition of temperature just proposed. But if a real gas such as methane were employed instead of helium this empirical method of defining $\theta$ would lead us to conclude that $\theta$ is \emph{not} a thermodynamic temperature. This is why, in the following, we suppose that $w\ne 0$. This seems at first to make the problem more complicated, but this is not so. At the end we may take the limit: $w\to 0$ and recover the usual perfect-gas law.

\paragraph{Non-zero weight. Mechanical average:}

The word ``average'' enters in this paper in two ways: as a mechanical average and as a thermal average. In the present paragraph we only consider mechanical averages. Physically, it is supposed that the cylinder ends have so much inertia that they do not respond to individual collisions. Let the corpuscle energy be denoted $E$. The maximum altitude $z_m$ that the corpuscle would reach in the absence of the piston is given by: $E=w\,z_m$. The mechanical average force exerted by the corpuscle on the cylinder bottom is twice the corpuscle momentum $p$ when it collides with the plane, times the number of collisions per unit time $1/\tau$, where $\tau$ denotes the motion period. When the corpuscle does not reach the piston, $z_m≤h$, we have $\tau=\tau(z_m)$. That is: $F_o=2p/\tau$. The corpuscle momentum $p$ is, according to the Hamilton equations (see Appendix \ref{hamilton}) given by $dp/dt=-w$. Thus: $p=p_o-w\,t$ where $p_o$ is a constant. If $w\ne 0$ we may set $t=0$ when $p=0$, that is, at the top of the corpuscle trajectory. The corpuscle momentum at time $t$ is then $p=-w\,t$, and the collision time is $t=\tau(z_m)/2$ (see Figure \ref{figure}). Thus $F_o=-\left(2w\tau(z_m)/2\right)(1/\tau(z_m))=-w$. When the corpuscle possesses enough energy to reach the piston the motion period becomes: $\tau(z_m)-\tau(z_m-h)$. Then the force $F_o$ experienced by the cylinder bottom is in general:
\begin{equation}\label{rcebis}
\begin{cases}
F_o(z_m)=-w &z_m≤h,\\
F_o(z_m)=-w\,\frac{\tau(z_m)}{\tau(z_m)-\tau(z_m-h)} \qquad &z_m>h.
\end{cases}
\end{equation}

The force $F$ experienced by the piston when the corpuscle energy is $w\, z_m$, on the other hand, is likewise:
\begin{equation}\label{orcebis}
\begin{cases}
F(z_m)=0 &z_m≤h,\\
F(z_m)=w\,\frac{\tau(z_m-h)}{\tau(z_m)-\tau(z_m-h)} \qquad &z_m>h.
\end{cases}
\end{equation}
so that irrespectively of $z_m$ we have: $F_o+F=-w$. This means that the cylinder, considered as a rigid object of negligible weight, has an effective weight precisely equal to $w$, a most intuitive result. By linearity, the same conclusion must hold for average forces: $\ave{F_o+F}=-w$, for any energy distribution. This will be shown explicitly below for the appropriate energy distribution $\om(z_m)$.

\paragraph{Non-zero weight. Thermal average:}

Because the cylinder lower end is in contact with a bath it suffers a slight thermal motion and the corpuscle energy $w\,z_m$ slowly varies in the course of time. Accordingly, the average force $\ave{F_o}$ experienced by the bottom of the cylinder and the average force $\ave{F}$ experienced by the piston are respectively, from \eqref{rcebis} and \eqref{orcebis}, if $\om(z_m)$ denotes the distribution:
\begin{align}\label{moh}
\ave{F_o}=\frac{\int_{0}^\infty dz_m\, \om(z_m)F_o(z_m)}{\int_0^{\infty}\, dz_m\,  \om(z_m)}
=-w \frac{\int_{0}^h dz_m\, \om(z_m)+\int_{h}^\infty dz_m\, \om(z_m)\frac{\tau(z_m)}{\tau(z_m)-\tau(z_m-h)}}{\int_0^{h}\, dz_m\, \om(z_m)+\int_{h}^\infty dz_m\, \om(z_m)}.
\end{align}
\begin{align}\label{mo}
\ave{F}=\frac{\int_{0}^\infty dz_m\, \om(z_m)F(z_m)}{\int_0^{\infty}\, dz_m\,  \om(z_m)}
=w \frac{\int_{h}^\infty dz_m\, \om(z_m)\frac{\tau(z_m-h)}{\tau(z_m)-\tau(z_m-h)}}{\int_0^{h}\, dz_m\, \om(z_m)+\int_{h}^\infty dz_m\, \om(z_m)}.
\end{align}

According to our simplicity principle, the average forces must be \emph{independent} of the corpuscle equation of motion, and thus of the $\tau(.)$-function. This condition obtains from \eqref{moh} and \eqref{mo} if one selects the following  distribution:
\begin{equation}\label{pond}
\begin{cases}
\om(z_m)=\exp(-w z_m/\theta)\tau(z_m) &z_m≤h,\\
\om(z_m)=\exp(-w z_m/\theta)\p\tau(z_m)-\tau(z_m-h)\q \qquad &z_m>h,
\end{cases}
\end{equation}
where $\theta$ has the dimension of an energy. The average forces become, using \eqref{moh}, \eqref{mo} and \eqref{pond}:
\begin{align}\label{moy}
\ave{F_o}&=-w\frac{\int_{0}^\infty dz_m\, \exp(-w z_m/\theta)\tau(z_m)}{\int_0^{h}\, dz_m\, \exp(-w z_m/\theta)\tau(z_m)+\int_{h}^\infty dz_m\, \exp(-w z_m/\theta)\p\tau(z_m)-\tau(z_m-h)\q}\nonumber\\
&=\frac{w}{\exp(-w\,h/\theta)-1}\to
\begin{cases}
 -\frac{\theta}{h}\quad w\,h\ll \theta\\
 -w\quad w\,h\gg \theta
\end{cases} 
 \nonumber\\
\ave{F}&=\frac{w \int_{h}^\infty dz_m\, \exp(-w z_m/\theta)\tau(z_m-h)}{\int_0^{h}\, dz_m\, \exp(-w z_m/\theta)\tau(z_m)+\int_{h}^\infty dz_m\, \exp(-w z_m/\theta)\p\tau(z_m)-\tau(z_m-h)\q}\nonumber\\
&=\frac{w}{\exp(w\,h/\theta)-1}\to 
\begin{cases}
 \frac{\theta}{h}\quad w\,h\ll \theta\\
 0\quad w\,h\gg \theta
\end{cases}
\end{align}
with: $\ave{F_o}+\ave{F}=-w$ since: $1/(\exp(-x)-1)+1/(\exp(x)-1)+1=0$: The cylinder weight is $w$ irrespectively of the temperature (leaving aside the weight of the cylinder walls). Similar to what was done in Section \ref{barometric}, in the above integrals going from $h$ to $\infty$ we have replaced $\exp(-w z_m/\theta)$ by $\exp(-w h/\theta)\exp(-w (z_m-h)/\theta)$ and introduced the variable $z_m'\equiv z_m-h$, so that all the integrals go from zero to infinity and cancel out. From now on we will omit the averaging signs on $\ave{F_o}$ and $\ave{F}$.

For a collection of $N$ independent corpuscles having weights $w_i,~i=1,...N$ respectively, the force is a sum of $N$  terms of the form given in \eqref{moy}. In the case of zero weights ($w_i$=0 or more precisely: $w_i\,h\ll \theta$), the above expression gives: $F\,h=\theta$. Thus we have obtained the perfect-gas law: $F\,h=N\,\theta$. The perfect-gas law does not depend on the nature of the corpuscles.

\paragraph{Average force for a three-dimensional space:}

We suppose that the cylinder radius is very large compared with $h$ and we do not consider the force exerted by the corpuscle on the cylinder wall. Motion of the corpuscle along directions perpendicular to $z$ (say, $x$ and $y$) \emph{does} affect the round-trip time function $\tau(\zeta)$. However, since the average force does not depend on this function, the ideal-gas law is unaffected. This is so for any physical system involving a single corpuscle provided the  physical laws are invariant under a $z$-translation (besides being static).

The internal energy, to be discussed in the following section, though, is incremented. One can prove that in the non-relativistic approximation and in the absence of gravity the internal energy is multiplied by 3. It would be incremented further by corpuscle rotation or vibration, not considered here. Using conventional methods, Landsberg \cite{landsberg} and Louis-Martinez \cite{Louis-Martinez:2011} obtain exactly the same result as given above (except for the factor 3 in the expression of the internal energy, relating to the number of space dimensions considered).

\section{Internal energy}\label{internal}

The gas internal energy $U$ is the average value of $E\equiv w\,z_m$, the gravitational energy being accounted for. Note that only corpuscule motion along the $z$-axis is being considered. For simplicity, we first assume here that the cylinder rests on the ground level: $h_o=0$. The expression of $U$ is, using the energy distribution given in \eqref{pond}:
\begin{align}\label{uuu}
U&=\frac{\int_0^{w\,h} dE\, E\exp(-E/\theta)\tau(E/w)+\int_{w\,h}^\infty dE\, E\exp(-E/\theta)\p\tau(E/w)-\tau(E/w-h)\q}{\int_0^{w\,h} dE\, \exp(-E/\theta)\tau(E/w)+\int_{w\,h}^\infty dE\, \exp(-E/\theta)\p\tau(E/w)-\tau(E/w-h)\q}\nonumber\\
&=\frac{\int_0^{\infty} dE\, E\exp(-E/\theta)\tau(E/w)-\int_{w\,h}^\infty dE\, (E-wh+wh)\exp(-E/\theta)\tau(E/w-h)}{\int_0^{\infty} dE\, \exp(-E/\theta)\tau(E/w)-\int_{w\,h}^\infty dE\, \exp(-E/\theta)\tau(E/w-h)}\nonumber\\
&=\frac{(1-\exp(-wh/\theta))\int_0^{\infty} dE\, E\exp(-E/\theta)\tau(E/w)-wh\int_{wh}^\infty dE\, \exp(-E/\theta)\tau(E/w-h)}{(1-\exp(-wh/\theta))\int_0^{\infty} dE\, \exp(-E/\theta)\tau(E/w)}\nonumber\\
&=\frac{(1-\exp(-wh/\theta))\int_0^{\infty} dE\, E\exp(-E/\theta)\tau(E/w)-wh\exp(-wh/\theta)\int_{0}^\infty dE\, \exp(-E/\theta)\tau(E/w)}{(1-\exp(-wh/\theta))\int_0^{\infty} dE\, \exp(-E/\theta)\tau(E/w)}\nonumber\\
&=\frac{\int_0^\infty dE\, E\exp(-E/\theta)\tau(E/w)}{\int_0^\infty dE\, \exp(-E/\theta)\tau(E/w)}-\frac{w\,h}{\exp(w\,h/\theta)-1}
\equiv U_1(\theta)+U_2(\theta,\,h).
\end{align}
If $h_o\ne 0$, one must add to $U$ the energy $w\,h_o$ required to raise the cylinder from the ground level to $h_o$, and $h=h_1-h_o$.

Going back to the expression of $U$ in \eqref{uuu} we note that the first term $U_1(\theta)$ minus $\theta$ corresponds to the kinetic energy $K$, while the second term $U_2(\theta,h)$, plus $\theta$, corresponds to the potential energy $P$. In the non-relativistic limit, the first term, minus $\theta$, gives the well-known expression $K=\theta/2\equiv \kB T/2$, see the proof at the end of the present section. Without gravity we have of course $P=0$. In the general case the splitting of $U$ into $K+P$ seems to be artificial. The internal energy $U(\theta,h)$ thus is the sum of a term function of $\theta$ but not of $h$ and a term which tends to $-\theta$ when $wh\ll \theta$. To evaluate the first term we need to know the round-trip time $\tau(z_m)$ to within an arbitrary proportionality factor and an integration must be performed in that case.

The expressions given earlier for the average force $\ave{F}$ in \eqref{moy} and the internal energy $U$ in \eqref{uuu} may be written, setting $\beta\equiv 1/\theta$, as:
\begin{align}\label{a}
\ave{F}&=\frac{\partial \ln(Z)}{\beta\, \partial h}\qquad U=-\frac{\partial \ln(Z)}{\partial \beta}\nonumber\\
Z(\beta,h)&=\p 1 - \exp(-\beta \,w\,h) \q \frac{w}{2\pi \hbar}\int_0^\infty dz_m\, \exp(-\beta \,w\,z_m)\tau(z_m).
\end{align}
$Z$ is the quantity called in statistical mechanics the partition function. The Planck constant $2\pi\hbar$ introduced to make $Z$ dimensionless plays no physical role in this paper. All the physical results may be derived from the above expression of $Z(\beta,h)$. 

\paragraph{Non-relativistic approximation}\label{non}

In the special case of non-relativistic motion\footnote{Let us recall the following mathematical result:\\ $\int_0^\infty dx\, \exp(-x)x^n=\Gamma(n+1)$, with 
$\Gamma (\frac{3}{2})=\frac{\sqrt{\pi}}{2}$, $\Gamma (n+1)=n!$ if $n$ is an integer, and $\Gamma(n+1)/\Gamma(n)=n$. Thus:  $\int_0^\infty dx\, \exp(-x)\sqrt{x}=\Gamma(\frac{3}{2})=\sqrt{\pi}/2$ and:  $\int_0^\infty dx\, \exp(-x)x\sqrt{x}=\Gamma(1+\frac{3}{2})=\frac{3}{2}\Gamma(\frac{3}{2})=\frac{3}{2}\frac{\sqrt{\pi}}{2}$, so that: $\frac{\int_0^\infty dx\, \exp(-x)x\sqrt{x}}{\int_0^\infty dx\, \exp(-x)\sqrt{x}}=\frac{3}{2}$.} we have, see Appendix \ref{hamilton}: $\tau(z_m)=2\sqrt{\frac{2m\,z_m}{w}}$ where $m$ denotes the corpuscle mass. Thus the first term in \eqref{uuu} is equal to $3\theta/2$ and the expression of the internal energy reads:
\begin{align}\label{ap}
U(\beta,h)=\theta\left(\frac{3}{2}-\frac{w\,h/\theta}{\exp(w\,h/\theta)-1}\right).
\end{align}

Let us give an example of application of the above formula. In a (Joule-Thomson) free expansion $U$ remains constant but the temperature may decrease. If the final cylinder height is infinite the new temperature $\tilde{\theta}=2U/3$ is given by:
\begin{align}\label{apbis}
\frac{\tilde{\theta}}{\theta}=1-\frac{2}{3}~\frac{w\,h/\theta}{\exp(w\,h/\theta)-1}.
\end{align}
The temperature may therefore be reduced up to the third of its initial value through unlimited free expansion. It is only for a perfect gas that the temperature remains constant.

When the non-relativistic approximation is made the expression of $Z$ given in \eqref{a} becomes:
\begin{align}\label{ukmb}
Z(\beta,h)&=\p 1 - \exp(-\beta \,w\,h) \q\frac{w}{2\pi \hbar}\int_0^\infty dz_m\, \exp(-\beta \,w\,z_m)2\sqrt{\frac{2m\,z_m}{w}} \nonumber\\
&\equiv \p 1 - \exp(-\beta \,w\,h) \q\frac{w}{2\pi \hbar}\times f(\beta) \nonumber\\ 
&=\p 1 - \exp(-\beta \,w\,h) \q \frac{w}{2\pi \hbar} 2\sqrt{\frac{2\,m}{w}}(\beta\,w)^{-3/2}\sqrt{\pi}/2\nonumber\\
&=\C \,\p 1 - \exp(-w\,h/\theta) \q \theta^{3/2},
\end{align}
where $\C$ is a constant ($w$ and $m$ are constants) that will not be needed in the following because we will only be interested in derivatives of $\ln(Z)$. For a perfect gas ($w\,h\to 0$) $Z=h\, \sqrt{\theta}$ to within an unimportant constant factor. 

\paragraph{Practical units:}

The energy $\theta=\ave{F}h$ has been defined so far only to within a multiplicative factor from dimensional considerations. This factor is fixed by agreeing that $\theta=273.16$ $\kB$ exactly when the cylinder is in thermal equilibrium with water at its triple point. Here $\kB= 1.38066...~  10^{-23}$ joules, is considered as an energy unit (akin to the calorie = 4.182... joules). This manner of defining $\theta$ is equivalent to the usual one, though expressed differently. The dimensionless quantity $T\equiv \theta/\kB$ is the usual unit of thermodynamic temperature, expressed in kelvin. From our viewpoint it would be better to convene that $\theta=1$ exactly as the hydrogen triple-point temperature (HTP). The value of $\theta$ at the water triple-point (WTP) for example would be obtained experimentally by measuring the efficiency of a reversible heat engine operating with WTP as a hot bath and HTP as a cold bath. The known value is: $\theta_{WTP}=19.737...$

Next, measurements have shown that the number of atoms in 0.012 kg of carbon 12 is: $N_A\approx 6.0221367~10^{23}$. For this quantity of matter called a mole, the ideal-gas law therefore reads: $\ave{F}h=N_A \theta$, or: $\textsf{P}\,\textsf{V}=R\,T$, with the ideal-gas constant: $R\equiv N_A \kB \approx 8.31451$ joules per kelvin per mole.

\section{Stability:}\label{temperaturestability}

Solutions obtained for the force $F$ and the energy $U$ imply stable equilibria provided two conditions be satisfied. Firstly, the isothermal compressibility $\kappa_T\equiv -(1/h)\partial h/\partial F$ must be positive. This is readily verified since the derivative of the force $F$ given in \eqref{moy} with respect to $h$ is negative. Secondly, one must verify that the isochore heat capacity $C$ is positive. This is a more difficult problem solved below. Given that $C$ and $\kappa_T$ are positive it follows that the isobaric heat capacity $c_p>C$ is positive, and the isentropic compressibility $\kappa_s=\kappa_T \,C/c_p$ is positive also. Thus, let us show that $C$ is positive.

\paragraph{$U$ is non-negative:}

Assuming that $\tau$ is derivable and $\tau(0)=0$ an integration by parts of the numerator of $U_1(\theta)$ in \eqref{uuu} gives 
\begin{align}\label{bpn}
\int_0^\infty dE\, \exp(-\beta E)\tau(E/w)=\frac{\phi(\beta)}{\beta},\qquad 
\phi(\beta)\equiv\frac 1 w\int_0^\infty dE\, \exp(-\beta E)\tau'(E/w). 
\end{align}
Since $\tau$ is non decreasing, $\phi(\beta)$ is non-negative and $\phi'(\beta)$ is negative. Thus 
\begin{align}\label{gbn}
U_1(\theta)=-\frac d{d\beta}\ln\frac{\phi(\beta)}{\beta}=\theta-\frac{\phi'(\beta)}{\phi(\beta)}>\theta. 
\end{align}
Since $U_2(\theta,h)>-\theta$, we have proven that: $U=U_1+U_2$ is positive.

\paragraph{$C$ is non-negative:} 

We have 
\begin{align}\label{bgbn}
\frac{\partial U_2(\theta,h)}{\partial \theta}&=-\bigg(\frac {\alpha/2}{\sinh (\alpha/2)}\bigg)^2>-1, \qquad \alpha\equiv \frac{wh}{\theta}, \nonumber\\
U_1'(\theta)&=1+\beta^2\frac {d^2}{d\beta^2}\ln\phi(\beta). 
\end{align}
In order to get $C>0$, it thus suffices to show that $\phi''\phi\geq\phi'^2$, that is, letting $f(\beta,E)\equiv\exp(-\beta E)\tau'(E/w)$ for short,  
\begin{align}\label{bbnk}
\bigg(\int_0^\infty dE\,E^2\,f(\beta, E)\bigg)\, \bigg(\int_0^\infty dE\,f(\beta, E)\, \bigg)\geq\bigg(\int_0^\infty dE\,E\,f(\beta, E)\, \bigg)^2.
\end{align}
Since $f$ is non-negative \eqref{bbnk} is the classical inequality regarding the moments of order 0, 1 and 2 of the mesure $\mu$ defined by: $d\mu=f(\beta,E)dE)$. Thus:
$C=\frac{\partial U}{\partial\theta}\ge 0$ and the expressions obtained from our simplicity principle imply stability of the equilibria.

\section{The Helmholtz fundamental relation.}

It is convenient to introduce the Helmholtz fundamental relation: $A(\theta,h)\equiv -\theta\ln(Z(\theta,\,h))$. The letter $A$ originates from the German “Arbeit” or work, but this letter may also stand for (constant temperature) \g{Available work}. The force $F_o$ that the corpuscle exerts on the base, the force $F$ that the corpuscle exerts on the piston and the internal energy result from the Helmhotz fundamental relation depending separately on $h_o$ and $h_1=h_o+h$. We consider thus a cylinder whose base has been raised from $z=0$ to $z=h_o$. The previous relations for $F_o,\,F$ in \eqref{moy} and for $U$ in \eqref{uuu} may be written as:
\begin{align}\label{uc}
F_o&=\frac{w}{\exp(-w\,h/\theta)-1}=-\frac{\partial A}{\partial h_o},\qquad F=\frac{w}{\exp(w\,h/\theta)-1}=-\frac{\partial A}{\partial h_1}\qquad \nonumber\\
U&=A-\theta\frac{\partial A}{\partial \theta}=\frac{\partial (\beta A)}{\partial \beta}\nonumber\\
A(\theta,h_0,h_1)&=-\theta\left( \ln(1-\exp(-w\,h/\theta))+\ln(\frac{w}{2\pi \hbar}\int_0^\infty dz_m\, \exp(-w\,z_m/\theta)\tau(z_m) \right)+w\,h_o
\end{align}
with $h\equiv h_1-h_o$. Thus, if the cylinder bottom is raised to an altitude $h_o$, $A$ and $U$ are both incremented by $w\,h_o$. From now on we set for simplicity $h_o=0$ unless specified otherwise.

We have obtained an expression for the Helmholtz fundamental relation $A(\theta,h)$ for the special case of a single corpuscle submitted to a constant force in the canonical ensemble. This fundamental relation has the same mathematical and physical content as the often-used energy fundamental relation: $U=U(S,h)$ and the entropy fundamental relation: $S=S(U,h)$, see \cite{callen}. The following expressions therefore coincide with the conventional ones applicable to any working substance.

\paragraph{The energy $\theta$ is a thermodynamic temperature}\label{thermo}

We prove in this section that $\theta$, introduced in previous sections on dimensional grounds only, \emph{is} a thermodynamic temperature. We do this by showing that the efficiency of a reversible thermal cycle employing ideal gases is: $1-\theta_c/\theta_h$, where $\theta_c$ is the cold-bath temperature and $\theta_h$ the hot bath temperature: this is the accepted Kelvin definition of absolute temperatures.

 From the law of conservation of energy the heat released by the gas is from \eqref{uc}:
\begin{align}\label{b}
-\delta Q\equiv dU+\ave{F} \,dh=dA-\frac{\partial A}{ \partial \theta}d\theta-\frac{\partial A}{ \partial h}dh-\theta \,d\p\frac{\partial A}{ \partial \theta}\q\equiv\theta \,dS,\qquad S=-\frac{\partial A}{ \partial \theta}.
\end{align}
For any function $f(\theta,h)$ such as $U,\,A,\,S$: $df\equiv \frac{\partial f}{ \partial \theta}d\theta+\frac{\partial f}{ \partial h}dh$. We employ only two independent variables namely $\theta$ and $h$ so-that partial derivatives are un-ambigous. If the gas is in contact with a thermal bath ($\theta$=constant) $\delta Q$ is the heat gained by the bath. The quantity $S$ defined above is called \g{entropy}. In particular, if heat cannot go through the gas container wall (adiabatic transformation) we have $\delta Q=0$ that is, according to the above result: $dS=0$. Thus reversible adiabatic transformations are isentropic. Note that $S$, here defined as the ratio of two energies, is dimensionless. It may therefore be written as the logarithm of a dimensionless quantity. The fact that $S$ defined above is a state function suffices to prove that $\theta$ is a thermodynamic temperature as shown below.

\paragraph{The Carnot cycle:}
A Carnot cycle consists of two isothermal transformations at temperatures $\theta_l$ and $\theta_h$, and two intermediate reversible adiabatic transformations ($dS=0$). After a complete cycle the entropy recovers its original value and therefore $dS_c+dS_h=0$. According to \eqref{b}: $-\delta Q_c=\theta_c\, dS_c$, $-\delta Q_h=\theta_h\, dS_h$ and therefore $\delta Q_c/\theta_c+\delta Q_h/\theta_h=0$. Energy conservation gives the work $\delta W$ performed over a cycle from: $ \delta W+\delta Q_c+\delta Q_h=0$. The cycle efficiency is defined as the ratio of $\delta W$ and the heating $-\delta Q_h$ supplied by the hot bath. We have therefore: $\eta\equiv \frac{\delta W}{-\delta Q_h}=\frac{\delta Q_h+\delta Q_c}{\delta Q_h}= 1-\frac{\theta_c}{\theta_h}$, from which we conclude that $\theta$ is the \g{thermodynamic temperature}. Since Kelvin time, thermodynamics temperatures are strictly defined from Carnot (or other reversible) cycles efficiency. In practice, temperatures may be measured by other means and employed in other circumstances.

We have implicitly assumed in the above discussion that the working medium (presently an ideal gas) has reached the bath temperature before being contacted with it. Otherwise, there would be at that time a jump in entropy, and the cycle would no longer be reversible. Given initial $\theta,\,h$ values, the temperature change $d\theta$ for an increment $dh$ in the isentropic regime ($dS=0$) follows from the relation: $d\theta=-\p \frac{\partial S/\partial h}{\partial S/\partial \theta}     \q dh$, where $S(\theta,h)$ may be expressed in terms of $Z(\theta,h)$, \eqref{a}, from the above expressions. The details will be omitted. It suffices to know that $\theta$ may be varied by varying $h$, in a calculable manner, in an isentropic transformation.

\section{Expressions of the entropy}\label{entropy}

In \eqref{b} we have expressed the state function $S$ in terms of the Helmholtz potential $A$
\begin{align}\label{bc}
S=-\frac{\partial A}{ \partial \theta}=\frac{\partial \,\left(\theta \ln(Z)\right)}{ \partial \theta}=\beta U+\ln(Z(\beta,h))
\end{align}
using for $U$ the expression given in \eqref{uc}.

\paragraph{Perfect gases:}\label{inr}

For a perfect gas ($w\to 0$, non-relativistic approximation) the expression of $Z$ in \eqref{ukmb} may be written as: 
\begin{align}\label{ab}
Z(\beta,h)&\propto h\, \theta^{1/2}, \qquad\ln\left(Z(\beta,h)\right)=       \ln\left(h\, \theta^{1/2}\right)    =-\beta \,A(\beta,h)\nonumber\\
F&=-\frac{\partial A}{\partial h}=-\frac{1}{\beta}\frac{\partial (\beta A)}{\partial h}=\frac{\theta}{h} \qquad U=\frac{\partial (\beta A)}{\partial \beta}=\frac{\theta}{2}
\end{align}
These are the usual expressions for the equation of state and the internal energy of a perfect gas. 

According to \eqref{bc} and \eqref{ab} the entropy is:
\begin{align}\label{uhc}
S=-\frac{\partial A}{ \partial \theta}=\frac{\partial \left(\theta\ln(h\,\theta^{1/2})\right)}{\partial \theta}= \ln(h\,\theta^{1/2})+\frac{1}{2}.
\end{align} 
Since $U=\theta/2$, the fundamental entropy relation reads:
\begin{align}\label{uhmc}
S(U,h)= \ln\left(h\,\sqrt{2U}\right)+\frac{1}{2} .
\end{align} 
We recover from this expression again: $\frac{1}{\theta}=\frac{\partial S}{ \partial U}=\frac{1}{ 2\,U}$, and: $\frac{F}{ \theta}=\frac{\partial S}{ \partial h}=\frac{1}{ h}$, that is $F\,h=\theta$.

\paragraph{Ideal gases:}\label{nri}

We consider a non-zero corpuscle weight $w$. The procedure is the same as in the previous paragraph. Let us collect previous results in \eqref{uuu} and \eqref{ap} for the internal energy $U$, in \eqref{bc} for $Z$ and \eqref{ukmb} for the entropy $S$, applicable to a cylinder of height $h$:
\begin{align}\label{bts}
U&=U_1(\theta)-\frac{w\,h/\theta} {\exp(w\,h)-1}\approx\theta\left( \frac{3} {2}-\frac{w\,h/\theta} {\exp(w\,h/\theta)-1}\right)  \nonumber\\
S&=U/\theta+\ln(Z)\approx\left( \frac{3} {2}-\frac{w\,h/\theta} {\exp(w\,h/\theta)-1}\right)+\ln\p 1 - \exp(-w\,h/\theta) \q +\frac{3}{2}\ln(\theta)
\end{align}
in the non-relativistic approximation, to within a constant that makes the quantities dimensionless if desired. The fact that the entropy tends to -$\infty$ when $w\to 0$ should not be a cause for concern because only entropy \emph{differences} are considered.

\paragraph{Another form of the entropy:}\label{otherentropy}

Recall from \eqref{bc} that:
\begin{align}\label{bz}
S\equiv-\frac{\partial A(\beta,h)}{ \partial \theta}=-\beta \frac{\partial \ln(Z(\beta,h))}{ \partial \beta}+\ln(Z(\beta,h)).
\end{align}
The successive trajectory actions are discretized with spacings equal to the Planck constant, and thus the corresponding energies are correspondingly discretized with subscripts $k$ and we suppose that different $k$ values correspond to different energies. In the present classical paper the Planck constant is allowed at the end to assume arbitrarily small values. 

Let the $Z(\beta,h)$ function be written as a sum of terms $\exp(-\beta \eps_k(h))$ instead of an integral. Then the entropy $S$ may be written as:
\begin{align}\label{cbt}
S=-\sum_{k=0}^\infty p_k(\beta, h)\,\ln(p_k(\beta,h)),\quad p_k(\beta,h)= \frac{\exp(-\beta \eps_k(h))}{Z(\beta,h)}, \quad Z(\beta,h)=\sum_{k=0}^\infty \exp(-\beta \eps_k(h)),
\end{align}
as one readily verifies by substituting the expression of $p_k$ into the expression of $S$. The above is a simple mathematical transformation. However, when there is some uncertainty concerning the presence of a corpuscle in the cylinder, it is useful to interpret the $p_k$ as independent probabilities. The entropy may not decrease when constraints are removed or restored inside a thermally isolated body\cite{Mello:2013}. Let us quote these authors: ``A common formulation of the law of increase of entropy states that in a process taking place in a completely isolated system the entropy of the final equilibrium state cannot be smaller than that of the initial equilibrium state. This statement does not specify that thermal isolation is all that is needed for its validity, with no need for mechanical isolation''.

Let us consider now two boxes labeled ``A'' and ``B'' and a single corpuscle. If the corpuscle is in box ``A'' the probability that the level $kA$ be occupied is denoted $p_{kA}$ with: $-\sum_{k=0}^\infty p_{kA}=1$. If the corpuscle is in box ``B'' the probability that the level $kB$ be occupied is denoted $p_{kB}$ with $-\sum_{k=0}^\infty p_{kB}=1$.  When the corpuscle is in box A with independent probability $P_A$ and in box B with probability $P_B=1-P_A$, the $p_{kA}$ should be multiplied by $P_A$ and the $p_{kB}$ should be multiplied by $P_B$. 
We therefore have for the entropy in that case:
\begin{align}\label{bt}
\tilde{S}&=-\sum_{k=0}^\infty p_{kA}\,P_A\ln(p_{kB}\,P_A)-\sum_{k=0}^\infty p_{kB}\,P_B\ln(p_{kB}\,P_B)=S_A\,P_A+S_B\,P_B+\Delta S\nonumber\\
S_A&=-\sum_{k=0}^\infty p_{kA}\ln(p_{kA}), \qquad S_B=-\sum_{k=0}^\infty p_{kB}\ln(p_{kB}), \qquad \Delta S=P_A\ln(1/P_A)+P_B\ln(1/P_B).
\end{align}
The additional term $\Delta S$ accounts for the fact that it is not known with certainty whether the corpuscle is in box A or in box B. When the two boxes are identical ($S_A=S_B\equiv S$) we have: $\tilde{S}=S+\Delta S$.

\section{Change in entropy upon removal and restoration of a separation}\label{separation}

The process  presently discussed is often considered in relation with the so-called \g{ Gibbs's paradox}, see for example\cite{Nagle:2010}. Let us give an adapted quotation of his paper: ``A new state is created by inserting a partition into the 2V volume. It is clear that there should be no change in entropy in that process. The initial state was prepared knowing precisely which distinguishable particles were in each subvolume, but there is no such knowledge about the final state, so from the information point of view it is clear that the initial state and the final state are not equivalent''. We consider a thermally-isolated vertical cylinder of height $2h$ (from $z=0$ to $z=2h$) separated by an impermeable wall at altitude $h$. The lower part of the cylinder is labeled ``A'' and the upper part is labeled ``B''. A corpuscle is introduced in part A. The separation is removed: The internal energy $U$ is unchanged since the system is adiabatic and no work has been performed. Because $U$ is an increasing function of temperature and a decreasing function of the cylinder height, doubling the height amounts to reducing the temperature to a value that we denote $\theta\equiv1/\beta$. Then we restore the separation and show on the basis of our formulas (derived from the simplicity principle) that the total internal energy remains the same and that the total entropy also remains the same, in agreement with the first and second laws of thermodynamics. We treat first the simple case of perfect gases (no weight, non-relativistic approximation) and then the case of ideal gases (non-zero weight $w$, arbitrary hamiltonian function).

\paragraph{Perfect gases:}

For a perfect gas the entropy is, omitting unimportant constants: $S_1=C\ln(\frac{U}{2})+\ln(h)$ since part B does not contain any corpuscle, with $U$ the internal energy and $C$ the constant-volume heat capacity. Next we remove the separation. This can be done without any work employed and thus without any change of $U$. The new expression of the entropy is therefore: $S_2=C\ln(\frac{U}{2})+\ln(2h)$ since $U$ is unchanged:
\begin{align}\label{uhkc}
S_2=S_1+\ln(2).
\end{align} 

Let us now suppose that the impermeable separation is restored. Again, this can be done without changing the internal energy $U$ and thus the temperature. If we knew that the corpuscle is in part ``A'', the entropy $S_3$ would be reduced to $S_1$. However, there is a probability $P_A=\frac{1}{2}$ that the corpuscle be in part ``A'' and a probability $P_B=\frac{1}{2}$ that the corpuscle be in part ``B''. Thus, according to \eqref{bt} the entropy is now:
\begin{align}\label{ukc}
S_3=S_1-P_A\ln(P_A)-P_B\ln(P_B)=S_1-\frac{1}{2}\ln(\frac{1}{2})-\frac{1}{2}\ln(\frac{1}{2})=S_1+\ln(2)=S_2.
\end{align}
Thus the entropy has not been reduced by restoring the separation. In the present model the entropy remains the same. Since we are dealing with a single corpuscle the problem of corpuscle distinguishability does not arise.

\paragraph{Ideal gases:}\label{nri}

Let us recall the general expressions given earlier for the internal energy $U$ in \eqref{ap} and for the entropy $S=\beta U+\ln(Z)$ in \eqref{ukmb} for a cylinder of height $h_o$ at a temperature reciprocal $\beta_o$:
\begin{align}\label{bts}
&\beta_o U(\beta_o,h_o)=\beta_o U_1(\beta_o)-\frac{\alpha_o} {\exp(\alpha_o)-1}\equiv e(\beta_o)-\frac{\alpha_o} {\exp(\alpha_o)-1}, \qquad \alpha_o\equiv \beta_o\,w \,h_o  \nonumber\\
&S(\beta_o,h_o)=\beta_o U(\beta_o,h_o)+\ln(Z(\beta_o,h_o)) \nonumber\\
&=\left( \beta_o U_1(\beta_o)-\frac{\alpha_o} {\exp(\alpha_o)-1}\right)+\ln\p 1 - \exp(-\alpha_o) \q +\ln(f(\beta_o))\nonumber\\
&\equiv-\frac{\alpha_o} {\exp(\alpha_o)-1}+\ln\p 1 - \exp(-\alpha_o) \q +g(\beta_o)
\end{align}
for some functions $e,f,g$ of $\beta_o$, to within constants that makes the quantities dimensionless if desired. The fact that the entropy tends to -$\infty$ when $w\to 0$ should not be a cause for concern because only entropy \emph{differences} are considered. We have omited the term $\frac{1} {2}$ in the expression of $S$.
 
In the process of removing the separation the internal energy $U$ remains the same but the cylinder height becomes $2h$, and thus the temperature reciprocal assumes a new value that we denote $\beta$. The parameter $\alpha_o$ in the above expression of the internal energy becomes $2\beta \,w\,h\equiv 2\alpha$, the value of $\alpha$ being given by the relation:
\begin{align}\label{bgts}
\beta U=e(\beta)-\frac{2\alpha} {\exp(2\alpha)-1}, \qquad \alpha\equiv \beta\,w \,h.
\end{align}
When the separation is restored there is a probability $P_A$ that the corpuscle be in part A and a probability $P_B$ that the corpuscle by in part B with:
\begin{align}\label{ujkch}
P_A=\frac{1}{1+\exp(-\alpha)}, \qquad  P_B=\frac{1}{1+\exp(\alpha)},          
\end{align}
In the present situation the two parts are intrinsically the same except that part B is being raised by $h$ with respect to part A and that accordingly a term $w\,h$ must be added to its internal energy. Accordingly the total energy reads:
\begin{align}\label{ujkc}
\beta \tilde{U}=e(\beta)-\frac{\alpha} {\exp(\alpha)-1}+\alpha P_B=e(\beta)-\frac{\alpha} {\exp(\alpha)-1}+\frac{\alpha} {\exp(\alpha)+1}=\beta U.          
\end{align}

Let us now evaluate the changes of entropy. Without the separation the entropy is given by \eqref{bts} with the values of $\beta$ and $\alpha$ given above: 
\begin{align}\label{bmts}
S=g(\beta)-\frac{2\alpha} {\exp(2\alpha)-1}+\ln\p 1 - \exp(-2\alpha) \q .
\end{align}

When the separation is put back in place, the two boxes being identical they have the same entropy, but one must take into account the fact that there is a probability $P_A$ that the corpuscle be in box ``A'' and a probability $P_B$ that the corpuscle be in box ``B'' and employ \eqref{bt}. Thus:
\begin{align}\label{bhmts}
\tilde{S}&=g(\beta)-\frac{\alpha} {\exp(\alpha)-1}+\ln\p 1 - \exp(-\alpha) \q+\Delta S \nonumber\\
\Delta S&=-P_A\ln(P_A)-P_B \ln(P_B)=\ln(1+\exp(\alpha))-\frac{\alpha}{1+\exp(-\alpha)}
\end{align} 

We have therefore:
\begin{align}\label{ts}
\tilde{S}-S&=-\frac{\alpha}{\exp(\alpha)-1}+\ln(1-\exp(-\alpha))+\ln(1+\exp(\alpha))-\frac{\alpha}{1+\exp(-\alpha))}\nonumber\\
&+\frac{2\alpha}{\exp(2\alpha)-1}-\ln(1-\exp(-2\alpha)=0.
\end{align}
Thus the entropy remains unchanged when the separation is restored, in agreement with the statement previously cited.

To summarize, if we consider an isolated system with separations, the initial entropy is defined as the sum of the separated parts. When the separations are removed the entropy of the combined system increases. But when the separations are restored the entropy remains the same because we no longer know precisely where the corpuscles are. It would thus be appropriate to employ the term ``missing information (MI)'' in place of the term ``entropy'', as a number of authors have suggested.

\section{The barometric law for arbitrary static potentials:}\label{barome}

We consider here weights depending on the $z$-coordinate in a stepwise manner, thereby approximating continuously varying weights. This is the case on earth since the weight is inversely proportional to the square of the distance $z+R$ from the earth center, where $R\equiv 1/w_o$ denotes the earth radius. Alternatively, one may say that the potential is: $\phi(z)\propto 1-\frac{1}{1+z/R}$ (in that example however problems of convergence arise: Planet atmospheres slowly evaporate).

Let $\tau_i(\zeta)$ denote the round-trip time corresponding to a distance $\zeta$ from the top of the trajectory when the weight is a constant $w_i$ with $\tau_i(0)=0$, and by convention $\tau_i(\zeta)=0$ if $\zeta<0$. If the function $\tau(\zeta)$ denotes the round-trip time for a unit weight, we have for a weight $w_i$: $\tau_i(\zeta)=\frac{\tau(w_i\,\zeta)}{w_i} $ according to Appendix \ref{hamilton}. In the following we will use only this relation and the law of conservation of energy along some corpuscle motion. 

We here consider weights that depend on the altitude $z$ by steps. Specifically, for mathematical convenience the weights are supposed to be: $1/\Delta_1$ for $0<z<\Delta_1\equiv z_1$, $1/\Delta_2$ for $\Delta_1<z<\Delta_1+\Delta_2\equiv z_2,\cdot \cdot \cdot$. It follows that the potential is incremented by 1 as one goes from one step to the next. The weight is a discontinuous function of $z$ but the potential is continuous. We are looking for the time spent by the corpuscle above $z_n$ for some integer $n$ as a function of the energy $E$. This time is obviously equal to zero if $E<n$ because the level $n$ is not reached. For higher energy values we have a sum of terms, each corresponding to the successive steps above $n$, that is:
\begin{align}\label{mdy}
0≤&E≤n & &\quad 0\nonumber\\
n≤&E≤n+1 & &\quad \tau_{n+1}(\frac{E-n}{w_{n+1}})=\Delta_{n+1} \tau(E-n)=\Delta_{n+1} d(E-n)\nonumber\\
n+1≤&E≤n+2 & &\quad \tau_{n+2}(\frac{E-n-1}{w_{n+2}})+\tau_{n+1}(\frac{E-n}{w_{n+1}})-\tau_{n+1}(\frac{E-n-1}{w_{n+1}}) \nonumber\\
& & &\qquad=\Delta_{n+2} \tau(E-n-1)+\Delta_{n+1} \tau(E-n)-\Delta_{n+1} \tau(E-n-1)\nonumber\\
& & &\qquad=\Delta_{n+1} d(E-n)+\Delta_{n+2} d(E-n-1)\nonumber\\
\dots\dots\dots&\dots\dots\dots &&\nonumber\\
n+i-1≤&E≤n+i & &\quad\sum_{k=1}^{i}\Delta_{n+k}\, d(E-n-k+1)
\end{align}
In the last expressions we have set: $\tau(E-i+1)-\tau(E-i)\equiv d(E-i+1)$, $i=1,2...$.

With an energy distribution $\exp(-E/\theta)\equiv x^E$, the average round-trip time at $z_n$ is:
\begin{align}\label{ge}
\ave{T_n}&=\int_n^{n+1} dE\, x^E \Delta_{n+1} d(E-n)\nonumber\\
&+\int_{n+1}^{n+2} dE\, x^E \p\Delta_{n+1} d(E-n)+\Delta_{n+2} d(E-n-1)\q\nonumber\\
&+.........\nonumber\\
&=(\Delta_{n+1}x^n+\Delta_{n+2}x^{n+1}+\Delta_{n+3}x^{n+2}+\cdot \cdot \cdot)\,I(x)\qquad I(x)\equiv \int_0^{\infty} dE\, x^E  \,d(E).
\end{align}

Thus the probability that the corpuscle be above $z_n$ is:
\begin{align}\label{cdy}
\frac{\ave{T_n}}{\ave{T_0}}=\frac{\Delta_{n+1}x^{n}+\Delta_{n+2}x^{n+1}+\Delta_{n+3}x^{n+2}+...}{\Delta_{1}+\Delta_{2}x+\Delta_{3}x^2+...},
\end{align}
If we set the $\Delta$'s as unity we may identify $z$ and $n$ and obtain: $x^n=\exp(-z/\theta)$, that is the usual barometric law for a constant weight.

This expression may be compared with the result obtained traditionally from the ratio of the integrals of the generalized Boltzmann factor: $\exp(-\phi(z)/\theta)$ from $z_n$ to $\infty$ and from $0$ to $\infty$. Since $\phi(z_n)=n$, we obtain by that method for $z_{n-1}<z<z_{n}:$
\begin{align}\label{phi}
\int_{z_{n-1}}^{z_n} dz\,\exp \p-\frac{1}{\theta}( \frac{z-z_n}{\Delta_n}+n) \q\propto \Delta_{n} b^{n-1},
\end{align}
the proportionality factor being a function of $\theta$ only. This leads to our result in \eqref{cdy}.

If the $\Delta_n$ are equal to 0 (corresponding to very large weights) except for $n=l$ and $n=h$, we find that the corpuscle can only be in the steps $l$ or $h$ of widths $\Delta_l\equiv g_l$, $\Delta_h\equiv g_h$ respectively. For simplicity we suppose that the \g{degeneracies} of these two levels are equal: $ g_l=g_h$. The ratio of the probability that the corpuscle be in $h$ and the probability that is be in $l$ is: $\exp\frac{\phi_l-\phi_h}{\theta}$. This is the generalized Boltzmann factor. In other words, if we consider only two levels with the same degeneracy $g$ we obtain that the ratio of the populations $\nu_l$, $\nu_h$ as given by the usual Boltzmann factor. 

\paragraph{Force and energy:}
Let us indicate how the expressions of the force exerted by the corpuscle on a piston located at $z=h$ and the internal energy $U$ may be obtained from the previous discussion. The barometric law established above: $\rho(z)\propto \exp(-\beta \phi(z))$ is normalized from the condition that there is one corpuscle between $z=0$ and $z=h$. The normalized expression for $\rho(z)$ provides us with the weight comprised between $z$ and $z+dz$. An integration from $z=h$ to $z=\infty$ gives the total weight in that range of $z$, and thus the force $F$ exerted by the corpuscle on the piston. We consider a large number of samples, each containing a single corpuscle located below $z=h$ according to our model. The above calculation makes sense, nevertheless, if we are looking for the general force law $F$ that reproduces the known barometric law in the case of a large number of weightless pistons. The internal energy is obtained likewise by evaluating the internal energy element between $z=h$ to $z+dz$ and integrating from $z=0$ to $z=h$.

\section{Conclusion}\label{conclusion}

Let us recall the concepts introduced in the present paper. One can imagine that after having introduced the notion of corpuscules moving in vacuum, Democritus observed the elastic bounces of a unit weight on a balance and defined the weight \g{impulse} from the motion period. Not knowing the nature of the motion, e.g. the Newton laws, he may have thought of introducing an energy distribution such that the average force $\ave{F}$ \emph{does not depend} on the law of motion. This, as we have seen, may be done. This distribution involves for dimensional reasons a quantity $\theta$ having the dimension of energy. Considering a thermal engine operating between two baths at temperatures $\theta_c,\,\theta_h$ one finds on the basis of the principles just stated that the maximum efficiency is: $1-\theta_c/\theta_h$. This allows us to call $\theta$ the thermodynamic temperature, defined only up to a constant factor, which may be fixed by convening, for example, that $\theta=1$ at the hydrogen triple point.

The present paper provides a first-principle proof of the ideal-gas law, including a possible effect of a corpuscle weight $w$ with no knowledge of the round-trip time function $\tau(\zeta)$ being required. Our thesis is that this law may be obtained on the sole basis of the Democritus model of corpuscles and vacuum. It is indeed unnecessary to specify the law $z(t;E)$ of corpuscle motion, where $E$ denotes the energy. Explicit expressions of the internal energy are obtained provided the round-trip time  $\tau(E/w)$ be known. One can further show that the ideal-gas internal energy depends only on temperature in the absence of gravity. The theory presented is strictly classical. As such, it does not depend on the numerical value of the universal constant $\hbar$. However, the introduction of this quantity is needed to make the results dimensionless. We have proven the stability of the gas from the equations obtained for the force $F$ and the internal energy $U$. We have also shown that, at least for ideal gases, the energy and the entropy (or, better, the missing information) are unchanged when a separation is introduced. The case of non-uniform weights $w(z)$ has been treated, based on the same concepts.

\appendix

\section{The Hamilton formalism}\label{hamilton}

We consider a hamiltonian of the form
\begin{align}\label{cfdy}
\mathcal{H}(p,z)=\mathcal{H}(p)+\Phi(z),
\end{align}
whose value can be considered as the corpuscle energy: $E=\mathcal{H}(p_o)+\Phi(z_o)$ for some initial values of $p$ and $z$. $\Phi(z)=w\,z$ is the potential energy, with $-w$ the force exerted on the corpuscle. The equations of motion are:
\begin{align}\label{cfy}
v\equiv \frac{dz}{dt}=\frac{d\mathcal{H}}{dp}\qquad \frac{dp}{dt}=-\frac{d\Phi(z)}{dz}=-w,
\end{align}

They are best understood in wave optics terms, see for example\cite{Arnaud:1973}, with $E=\hbar\om,~p=\hbar k$. The quantity analogous to energy is the wave angular frequency $\om\equiv2\pi$ divided by the time period, and the quantity analogous to momentum $p$ is the wave number ($2\pi$ divided by the space period), that is: $\hbar\om=\mathcal{H}(\hbar k)+\Phi(z)$. The first relation in \eqref{cfy} says that the group velocity $v$ is the derivative of $\om(k)$ with respect to $k$, as one can easily see by tracing two arrays of parallel lines on a plane with slightly different time and space periods. The second equation follows from the fact the $\om$ does not depend on time when time does not appear explicitly in the expression of $\mathcal{H}$ as is presently the case. As an example, set $w=0$ and consider the dispersion equation for gravity waves (waves on deep sea): $\om=\sqrt{g\,k}$ where $g$ denotes the gravity acceleration on earth. Then: $v=\frac{1}{2}\sqrt{\frac{g}{k}}$: the greater the wavelength the faster is the wave packet moving. The theory presented in the main text is applicable without any change to gravity waves. It provides in principle the average force exerted by such waved on two parallel reflectors some distance apart and at some temperature (this example is somewhat academic because at ordinary temperatures that force is negligible). Let us now go back to the mechanical problem.

The solution of the second equation in \eqref{cfy} is: $p=p_o-w\,t$. Provided $w\ne 0$ we may select a time origin such that this relation reads: $p=-w\,t$. This is assumed henceforth.

\paragraph{The non relativistic approximation:}\label{nra}

Through a second-order expansion we take $\mathcal{H}(p)$ to be of the form: $a_o+a_1\,p+a_2\,p^2$ where $a_o,a_1,a_2$ are constants, and obtain:
\begin{align}\label{cfhdy}
\frac{dz}{dt}=a_1+2a_2\,p=a_1-2a_2\,w\,t,\qquad z=z_o+a_1\,t-a_2\,w\,t^2,
\end{align}
with $z_o$ a constant. The energy is equal to $w\,z_m$ where $z_m$ is the maximum altitude reached by the corpuscle at time $t_m=\frac{a_1}{2a_2\,w}$, hence $z_m=z_o+\frac{a_1^2}{4a_2\,w}$. The round-trip time $\tau=t_2-t_1$ is the difference of the successive times at which $z=0$, that is $t_{1,2}$ are the two solutions of: $a_2\,w\,t^2-a_1\,t+\frac{a_1^2}{2a_2\,w}-z_m=0$. For $a_1=0$ we obtain: $\tau=2\sqrt{\frac{z_m}{a_2\,w}}=2\sqrt{\frac{2m\,z_m}{w}}$ if we set: $a_2\equiv \frac{1}{2m}$. The same result is obtained for any value of $a_1$. The non-relativistic approximation is thus a straightforward application of the second-order expansion of any $\mathcal{H}(p)$ function.

When $w$ is a constant the solution of the Hamilton equations is: $w(z-z_o)=-\mathcal{H}(-w\,(t-t_o))$ as one can see by taking the derivative of that expression of $z$ with respect to time and $z_o,\,t_o$ are integration constants, since $p=p_o-w\,t$ as seen above. A change of the corpuscle energy $E$ with respect to the $z=0$ level corresponds to a change of the $z_o$ constant the maximum altitude reached $z_m$ being equal to $E/w$. Omitting these constants, it follows from the previous relations that $\tilde{z}\equiv w\,z$ is a function of $\tilde{t}\equiv w\,t$ that does not depend on $w$. In the above discussion the function $\mathcal{H}(p)$ is supposed to have been selected once for all but is otherwise arbitrary.  

It follows that if we consider two weight values, say: $w_1$ and $w_2$ letting the subscripts 1 and 2 refer to the two weight values, if $w_1\,t_1=w_2\,t_2$, then $w_1\,z_1=w_2\,z_2$. Accordingly, if the function $\tau(z_m)$ is known for the case where $w=1$, the function $\tau(z_m; w)$ for an arbitrary weight $w$ is given by $\tau(z_m;w)=\frac{\tau(w\,z_m)}{w}$. For the case where $\mathcal{H}(p)=p^2/2$ for example, we have seen that: $\tau(z_m;w)=2\sqrt{\frac{2m\,z_m}{w}}$. The relation $\tau(z_m;w)=\frac{\tau(w\,z_m)}{w}$ is indeed satisfied with $w_1=1,~w_2=w$.

\bibliographystyle{unsrt}
\bibliography{ideal}

\end{document}